\documentclass[twocolumn,english,showpacs,preprintnumbers,amsmath,amssymb,floatfix]{revtex4-1}

\usepackage[T1]{fontenc}
\usepackage[latin9]{inputenc}
\usepackage{color}
\usepackage{array}
\usepackage{amstext}
\usepackage{graphicx}
\usepackage{esint}
\usepackage{rotating}
\usepackage[font=small,labelfont=bf]{caption}
\usepackage{xcolor}
\usepackage[colorlinks = true,
linkcolor = magenta,
urlcolor  = blue,
citecolor = red,
anchorcolor = blue]{hyperref}

\makeatletter

\@ifundefined{textcolor}{}
{%
 \definecolor{BLACK}{gray}{0}
 \definecolor{WHITE}{gray}{1}
 \definecolor{RED}{rgb}{1,0,0}
 \definecolor{GREEN}{rgb}{0,1,0}
 \definecolor{BLUE}{rgb}{0,0,1}
 \definecolor{CYAN}{cmyk}{1,0,0,0}
 \definecolor{MAGENTA}{cmyk}{0,1,0,0}
 \definecolor{YELLOW}{cmyk}{0,0,1,0}
 }

\@ifundefined{definecolor}
 {\usepackage{color}}{}
\@ifundefined{definecolor}
 {\usepackage{color}}{}
\makeatother
\usepackage{babel}

\begin{document}


\title {EMC effect in the next-to-leading order approximation based on the Laplace transformation}

\author {Javad Sheibani$^{1}$}
\email{J.Sheibani@stu.yazd.ac.ir}

\author{ Abolfazl Mirjalili$^{1}$}
\email{A.Mirjalili@yazd.ac.ir (Corresponding Author)}

\author {S. Atashbar Tehrani$^{2}$}
\email{Atashbart@gmail.com}

\affiliation {
$^{(1)}$Physics Department, Yazd University, P.O.Box 89195-741, Yazd, Iran       \\
$^{(2)}$Independent researcher, P.O.Box 1149-834413, Tehran, Iran    }

\date{\today}

%
\begin{abstract}\label{abstract}
{{} In this article, using Laplace transformation , an analytical solution is obtained
for the DGLAP evolution equation at the next-to-leading order of perturbative QCD.
The technique is also employed to extract, in the Laplace $s$-space, {{} an} analytical solution
for the nuclear structure function, $F_2^A(x, Q^2)$.} {{}Firstly,}  the results for  separate
nuclear parton distributions for all parton types are presented which  include valence quark densities,
the anti-quark and strange sea PDFs and finally the gluon distribution.
Based on the Laplace transformation, the obtained parton distribution functions  and the nuclear
 structure function in the $x$-space  are  compared with the results from the
 {\tt AT12} \href{http://dx.doi.org/10.1103/PhysRevC.86.064301}{{\rm Phys.\ Rev.\ C} {\bfseries 86}, 064301 (2012)} model.  Our calculations are in  good agreement with the available
 DIS experimental data  as well as theoretical models  which contain both small and large
 values of $x$-Bjorken variable.
{{} We compare our nuclear PDFs sets with those from other recent
collaborations, in particular with the {\tt nCTEQ15} and {\tt HKN07} sets.} The comparison between our results and those from the literature indicates a good agreement .
\end{abstract}

\maketitle

%
%
\section{Introduction}\label{Introduction}
{{}QCD factorization theorems~\cite{Collins:1985ue,Bodwin:1984hc,Collins:1998rz} and
 parton distribution functions (PDFs), create a framework to fully describe nucleons.
A wide range  of different hard scattering processes, including  deep inelastic scattering (DIS),
Drell-Yan (DY) lepton pair production, vector boson production and the inclusive jet production
can be employed to determine PDFs through a global analysis.}
Considering the parton distribution functions inside nuclei,
characterized by the {{}atomic mass  and atomic number,} $A$ and $Z$ respectively,
{{}it is possible to achieve a proper theoretical description of hard scattering processes
which occurs in lepton-nucleon and proton-nucleon interactions.}
{{}The nucleon bound states can be described by nuclear PDFs (nPDFs)
and finally the nucleus can be  parameterized effectively in terms of the bonded nucleons.}
{{}Strong interactions between the nucleons in a nucleus were first recognized as
EMC effects which can theoretically be described by the exchange quark model \cite{hod1,hod2}.}
These interactions are characterized by the nPDFs {{}and will affect the bounded
nucleon structure.} Like the PDFs of free nucleons, the nPDFs can be obtained
by {{}fitting experimental data for nuclear deep inelastic scattering as well as
nuclear collisions.}

To access the PDFs and then nPDFs, it is required to get the solution of Dokshitzer-Gribov-Lipatov-Altarelli-Parisi (DGLAP) evolution equations ~\cite{Dokshitzer:1977sg,Gribov:1972ri,Lipatov:1974qm,Altarelli:1977zs}.DGLAP using the Laplace transform technique, some analytical solutions of these equations have been
reported in recent years ~\cite{Block:2010du,Block:2011xb,Block:2010fk,Block:2009en,Block:2010ti,Zarei:2015jvh,Taghavi-Shahri:2016ktz,Boroun:2015cta,Boroun:2014dka} ,which {{}have resulted in}
noticeable  success from the phenomenological point of view. {{}There has also been some
progress toward extracting the analytical solutions of the proton spin-independent structure function
$F_2^p (x, Q^2)$~\cite{Khanpour:2016uxh}, charged-current structure functions $xF_3(x, Q^2)$~
\cite{MoosaviNejad:2016ebo}, and also the spin-dependent one, i.e., $x g_1^p (x, Q^2)$, at the next-to-leading order (NLO) and
next-to-NLO (NNLO)  approximations ~\cite{AtashbarTehrani:2013qea,Salajegheh:2018hfs},
using the  Laplace transform technique.}

{{}In this paper, the required analysis has been performed, using sequential Laplace transforms
which lead us to an analytical solution of the DGLAP evolution equations at NLO
approximation. For this purpose singlet, non-singlet, and  individual gluon distributions inside the nucleus
are analytically calculated.} We present our results for the valence quark distributions $x u_v^A$ and $x d_v^A$, the anti-quark distributions $x \overline{u}^A $ and  $x\overline{d}^A$, the strange sea distribution $x \overline{s}^A$, and finally for the gluon distribution $x g^A$ inside the nucleus.
{{}Furthermore, we extract the analytical solutions for the nuclear structure function
$F_2^A(x, Q^2)$ as the sum of a flavor singlet $F_2^{\rm S}(x, Q^2)$, gluon $F_2^{\rm g}(x, Q^2)$,
and a flavor non-singlet $F_2^{\rm NS}(x, Q^2)$. The obtained results indicate an excellent agreement
with the DIS data as well as those obtained by other methods such as the fit to $F_2^{A^{\prime}}/F_2^A$ structure function ratio performed by the {\tt AT12} model~\cite{AtashbarTehrani:2012xh}.}

{{}The remainder of this paper  consists of the following sections:}
In Sec.~\ref{Sec2:Theoretical-formalism}, we shall provide a brief discussion {{}on}
the theoretical formalism to obtain the PDFs at the NLO approximation in perturbative QCD, based on the
Laplace transformation technique . In Sec.~\ref{Sec3:emc-effect}. we discuss the
theoretical formalism of the EMC effect and how to parametrize the nPDFs at the initial input scale.
In Sec.~\ref{Sec4:Nuclear-structure-function}, {{}details of extracting nuclear structure function
$F_2^A(x, Q^2)$ in Laplace space would be discussed.} Sec.~\ref{Sec5:Results} is devoted to present our
results, based on the  Laplace transformation. Finally, we give our summary and conclusion in
Sec.~\ref{Sec6:Summary}.

\section{A brief review on the solution of  DGLAP evolution equations, using the  Laplace transform technique}\label{Sec2:Theoretical-formalism}
The singlet $x q_{\rm S}(x,Q^2)$ and gluon $x g(x,Q^2)$ distribution functions
can be described by the DGLAP evolution equations  ~\cite{Dokshitzer:1977sg,Gribov:1972ri,Lipatov:1974qm,Altarelli:1977zs}. At the next-to-leading order approximation, {{}in the convolution notation} $\otimes$, the coupled DGLAP evolution equations  can be written as~\cite{Block:2007pg,Block:2008xc}
\begin{widetext}
\begin{eqnarray}\label{eq:singletF}
	&& \frac{4\pi}{\alpha_s(Q^2)} \frac{\partial F_{\rm S}} {\partial lnQ^2}(x,Q^2)
	= F_{\rm S}\otimes\left(P_{qq}^0 + \frac{\alpha_s(Q^2)} {4\pi}P_{qq}^1\right)(x,Q^2)
	 +\ G \otimes \left(P_{qg}^0 + \frac{\alpha_s(Q^2)}{4\pi} P_{qg}^1\right)(x,Q^2)\, ,
\end{eqnarray}
\begin{eqnarray}\label{eq:singletG}
	&& \frac{4\pi}{\alpha_s(Q^2)}\frac{\partial G}{\partial lnQ^2}(x,Q^2)
	= F_{\rm S}\otimes\left(P_{gq}^0 + \frac{\alpha_s(Q^2)} {4\pi}P_{gq}^1\right)(x,Q^2)
	 +\ G\otimes\left(P_{gg}^0 + \frac{\alpha_s(Q^2)} {4\pi} P_{gg}^1\right)(x,Q^2)\;. \label{s-g}
\end{eqnarray}
{{}In Eq.(\ref{s-g}), $\alpha_s(Q^2)$ is the running coupling constant and the Altarelli-Parisi
splitting kernels with one and two-loop corrections are denoted respectively by
$P_{ij}^0(x,\alpha_s(Q^2))$ and $P_{ij}^1(x,\alpha_s(Q^2))$ ~\cite{Altarelli:1977zs,Curci:1980uw,Furmanski:1980cm}.}
{{}The masses of charm, bottom, and top quarks ($m_c$,$m_b$, $ m_t$) would be taken into account in the energy scale $\mu$ by setting the number of active quark flavors; for $m^2_c < \mu^2 < m^2_b$ we would set N$_f$ = 4, and for $m^2_b < \mu^2 < m^2_t$  we would fix  N$_f$ = 5 in the evolution equations.} {{}Through this}, the QCD parameter $\Lambda$  can be adjusted at each heavy quark mass threshold, $\mu^2 = m^2_c$ and $m^2_b$. Therefore, when $N_f$ changes at $c$ and $b$ mass thresholds, the renormalized coupling constant $\alpha_s(Q^2)$ {{} will be}  continuously running ~\cite{Botje:2010ay}.

It is now possible to discuss briefly the method which is based on the Laplace transformation technique to
extract  analytical solutions for the  parton distribution functions, using the DGLAP evolution equations.
The evolution equations presented in Eqs.(\ref{eq:singletF},and~\ref{eq:singletG}) can be rewritten
with respect to $\nu$ and $\tau$ variables and in term  of the convolution integrals where
$\nu \equiv \ln (1/x)$ and $\tau$ is defined as $\tau(Q^2, Q_0^2) \equiv {1 \over 4\pi} \int_{Q^2_0}^{Q^2} \alpha_s({Q^{\prime}}^2) d \ \ln {Q^{\prime}}^2$ \cite{Block:2010du,Khanpour:2016uxh}.
Consequently, the related DGLAP equations are appeared as ~\cite{Block:2010du,Block:2011xb}
\begin{eqnarray}\label{eq:flaplacepartial1sspace}
	{\partial f\over \partial \tau }(s,\tau) & = & \left( \Phi_f^{\rm LO}(s) \ + \ \frac{\alpha_s(\tau
		)}{4\pi} \Phi_f^{\rm NLO}(s)\right) f(s,\tau)
	+  \left(\Theta_f^{\rm LO}(s) \ + \ \frac{\alpha_s(\tau)}{4\pi} \Theta_f^{\rm NLO}(s)\right)g(s,\tau) \;,
\end{eqnarray}
\begin{eqnarray}\label{2sspace}
	{\partial g\over \partial \tau }(s,\tau) & = &\left( \Phi_
	g^{\rm LO}(s) \ + \ \frac{\alpha_s(\tau)}{4\pi} \Phi_g^{\rm NLO}(s)\right) g(s,\tau)
	 + \left(\Theta_g^{\rm LO}(s) \ + \ \frac{\alpha_s(\tau)}{4\pi} \Theta_g^{\rm NLO}(s)\right)f(s, \tau)\;.
\end{eqnarray}
It should be noted that in deriving the above equations, the following property is used. The Laplace
transform of convolution factors is simply the ordinary product of the Laplace transform of the factors.
This is the reason {{}why} the usual  DGLAP equation can be converted to ordinary first order
differential equation in Laplace space $s$ with respect to $\tau$ variable as in Eqs.
(\ref{eq:flaplacepartial1sspace},and~\ref{2sspace}).

In continuation, the leading-order splitting functions of the PDFs, presented in Ref.~\cite{Altarelli:1977zs,Floratos:1981hs} in Mellin-space, are given {{}in} Laplace $s$ space by $\Phi_{(f,g)}^{\rm{LO}}$ and $\Theta_{(f,g)}^{\rm{LO}}$ ~\cite{Khanpour:2016uxh}:
\begin{eqnarray}
	\Phi_f^{\rm{LO}} = 4 - \frac{8}{3}\left(\frac{1}{s + 1} + \frac{1}{s + 2} + 2\left(\gamma_E + \psi (s + 1)\right)\right)\,,
\end{eqnarray}
\begin{eqnarray}
	\Theta_f^{\rm{LO}} = 2 N_f \left(\frac{1}{1 + s} - \frac{2}{2 + s} + \frac{2}{3 + s}\right)\,,
\end{eqnarray}
\begin{eqnarray}
	\Phi_g^{\rm{LO}} &=& 12 \left(\frac{1}{s} - \frac{2 }{1 + s} + \frac{1 }{2 + s} - \frac{1 }{3 + s} - \left(\gamma_E + \psi (s + 1)\right)\right)+ \frac{33 - 2 N_f}{3}\,,
\end{eqnarray}
\end{widetext}

and
\begin{eqnarray}
	\Theta_g^{\rm{LO}} = \frac{8}{3} \left(\frac{2}{s} - \frac{2}{1 + s} + \frac{1}{2 + s}\right)\,,
\end{eqnarray}
where $N_f$ is the number of active quark flavours, $\gamma _E$ is the Euler's constant and  $\psi$ is
the digamma function.

The next-to-leading order splitting functions  $\Phi_{(f,g)}^{\rm NLO}$ and  $\Theta_{(f,g)}^{\rm NLO}$
{{}have too long expressions} to be included here and {{}were} presented in
Appendix~{\bf A} of Ref.~\cite{Khanpour:2016uxh}.
A very simple parametrization  can be {{}taken} for
$\frac{\alpha_s(\tau)}{4 \pi}= a(\tau)$ as $a(\tau) = a_0$ .
One can {{}consider using the following expression for $a(\tau)$ in a generally more precise
calculation at the next-to-leading order (NLO) approximation, as in Ref.~\cite{Khanpour:2016uxh}.}
\begin{eqnarray}\label{eq:atau}
	a(\tau)\approx a_0+ a_1e^{-b_1\tau}\,.
\end{eqnarray}
{{}An excellent result accurate to a few parts in $10^4$ is obtained by this expansion.}
 Based on the definition of $a(\tau)$, given by {{}the} above equation, the following simplified
 notations for the splitting functions in $s$ space at the NLO approximation can be introduced where the
 conventions {{}presented} in Refs.~\cite{Block:2010du,Khanpour:2016uxh,Block:2011xb} are also
 used:

\begin{eqnarray}\label{eq:spliLOandNLO}
	\Phi_{f,g}(s)\equiv \Phi_{f,g}^{\rm LO}(s) + a_0 \Phi_{f,g}^{\rm NLO}(s), \nonumber \\
	\Theta_{f,g}(s)\equiv \Theta_{f,g}^{\rm LO}(s) + a_0 \Theta_{f,g}^{\rm NLO}(s)\;.
\end{eqnarray}
The solution of the coupled ordinary first order differential equations in Eqs.
(\ref{eq:flaplacepartial1sspace},and~\ref{2sspace}) at  the next-to-leading-order approximation
and  in terms of the initial distributions are straightforward. The evolved solutions
in the Laplace $s$ space at input scale $Q_0^2 = 2$ GeV$^2$, {{}taking into account
the initial distributions for the gluon $g^0(s)$ and singlet distributions $f^0(s)$,  are given
by~\cite{Block:2010du,Khanpour:2016uxh,Block:2011xb}:}
\begin{eqnarray}\label{eq:NLODGLAPSspace}
	f(s,\tau) & = & k_{ff}(a_1, b_1, s, \tau) \, f^0(s) + k_{fg}(a_1, b_1, s, \tau) \, g^0(s)\,, \nonumber \\
	g(s,\tau) & = & k_{gg}(a_1, b_1, s, \tau) \, g^0(s) + k_{gf}(a_1, b_1, s, \tau) \, f^0(s)\,. \nonumber \\
\end{eqnarray}
The analytical expressions {{}for the} coefficients
$k_{ff}$, $k_{fg}$, $k_{gf}$, and $ k_{gg}$  at the NLO approximation are given
in Appendix~{\bf B} of Ref.~\cite{Khanpour:2016uxh}.
\begin{figure}[htb]
	\begin{center}
		\includegraphics[clip,width=0.45\textwidth]{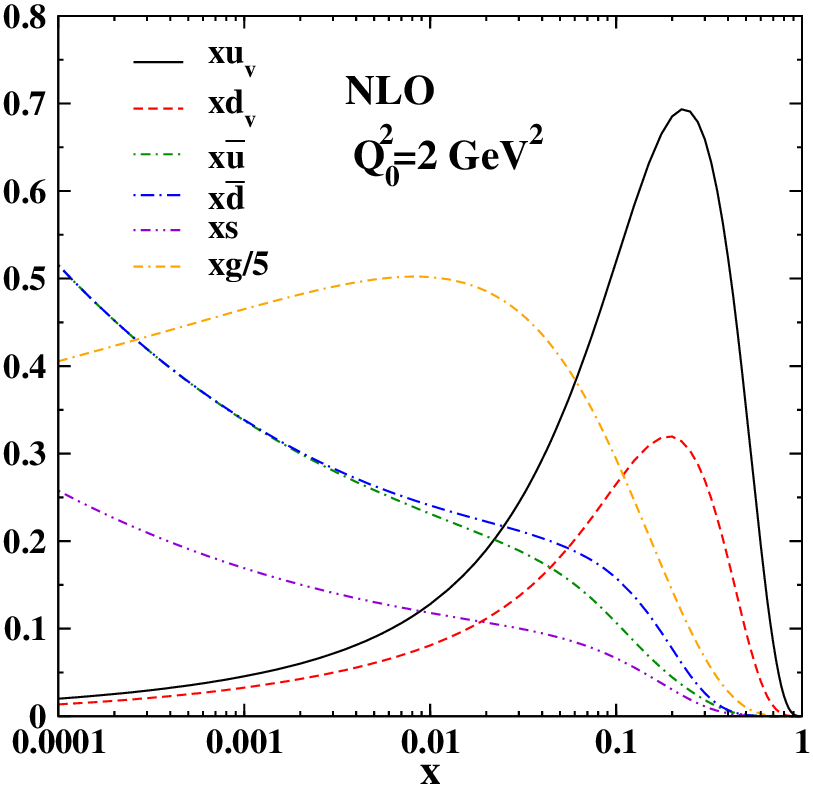}
		\caption{(Color online) Parton distribution for free proton at $Q_0^2=2$\; GeV$^2$.}\label{fig1:freepartonproton}
	\end{center}
\end{figure}
\begin{figure}[htb]
	\begin{center}
		\includegraphics[clip,width=0.45\textwidth]{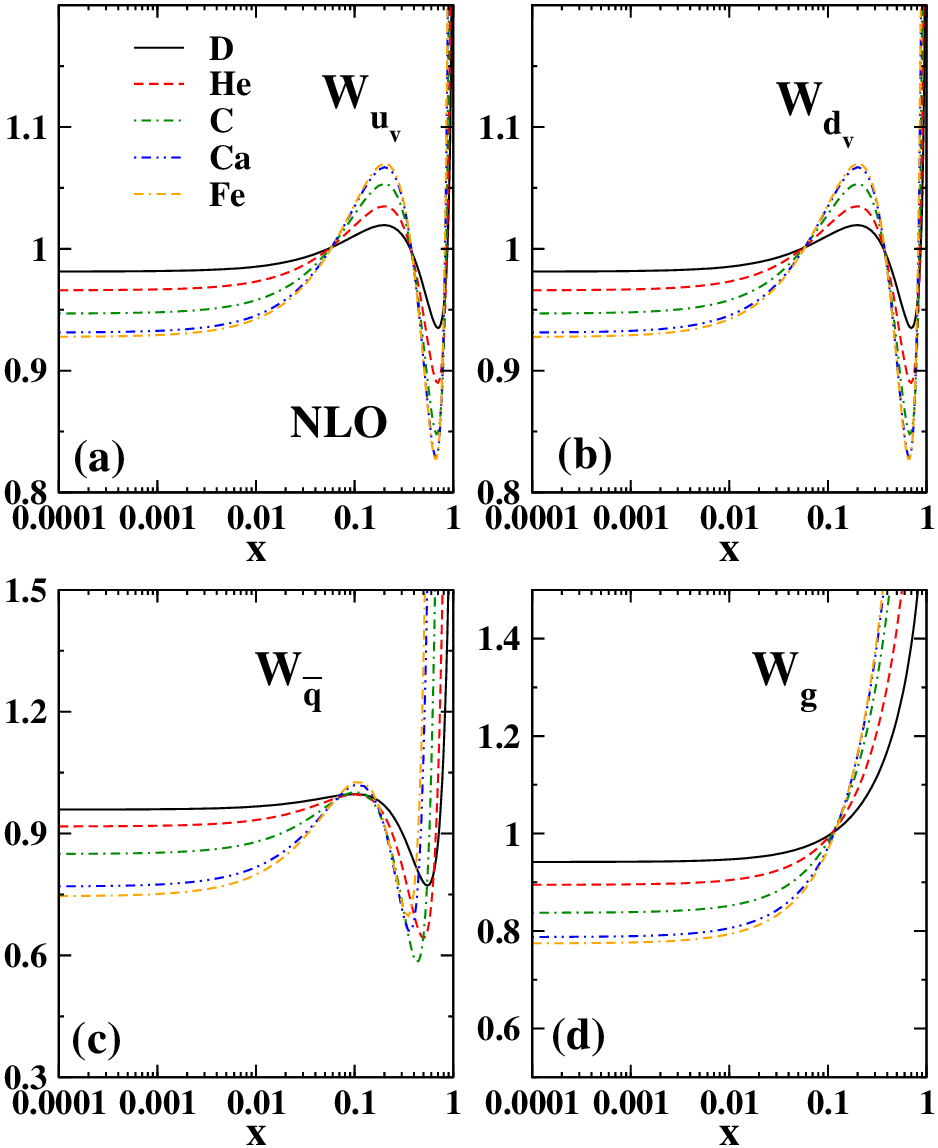}
		\caption{(Color online) Weight function for Fe, Ca, C, He and D nuclei, sub-leveled by (a) to (d) for different types of PDFs .}\label{fig2:weightfunction}
	\end{center}
\end{figure}
\begin{figure}[htb]
	\begin{center}
		\includegraphics[clip,width=0.45\textwidth]{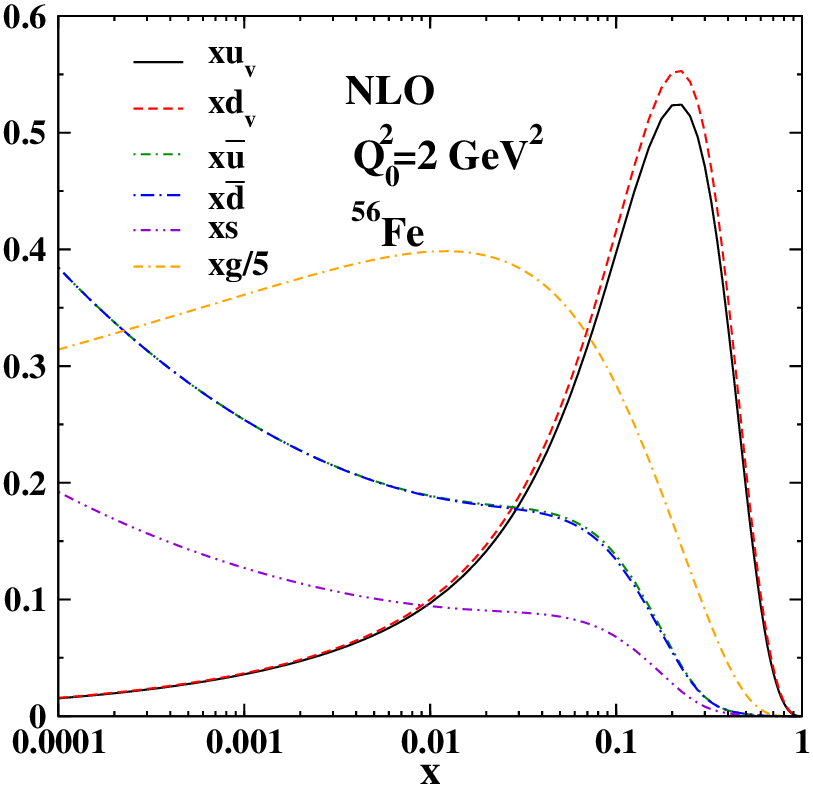}
		\caption{(Color online) Parton distribution for $^{56}Fe$ at $Q_0^2=2$\; GeV$^2$.}\label{fig3:FeQ2}
	\end{center}
\end{figure}
\begin{figure}[htbp]
	\begin{center}
		\includegraphics[clip,width=0.45\textwidth]{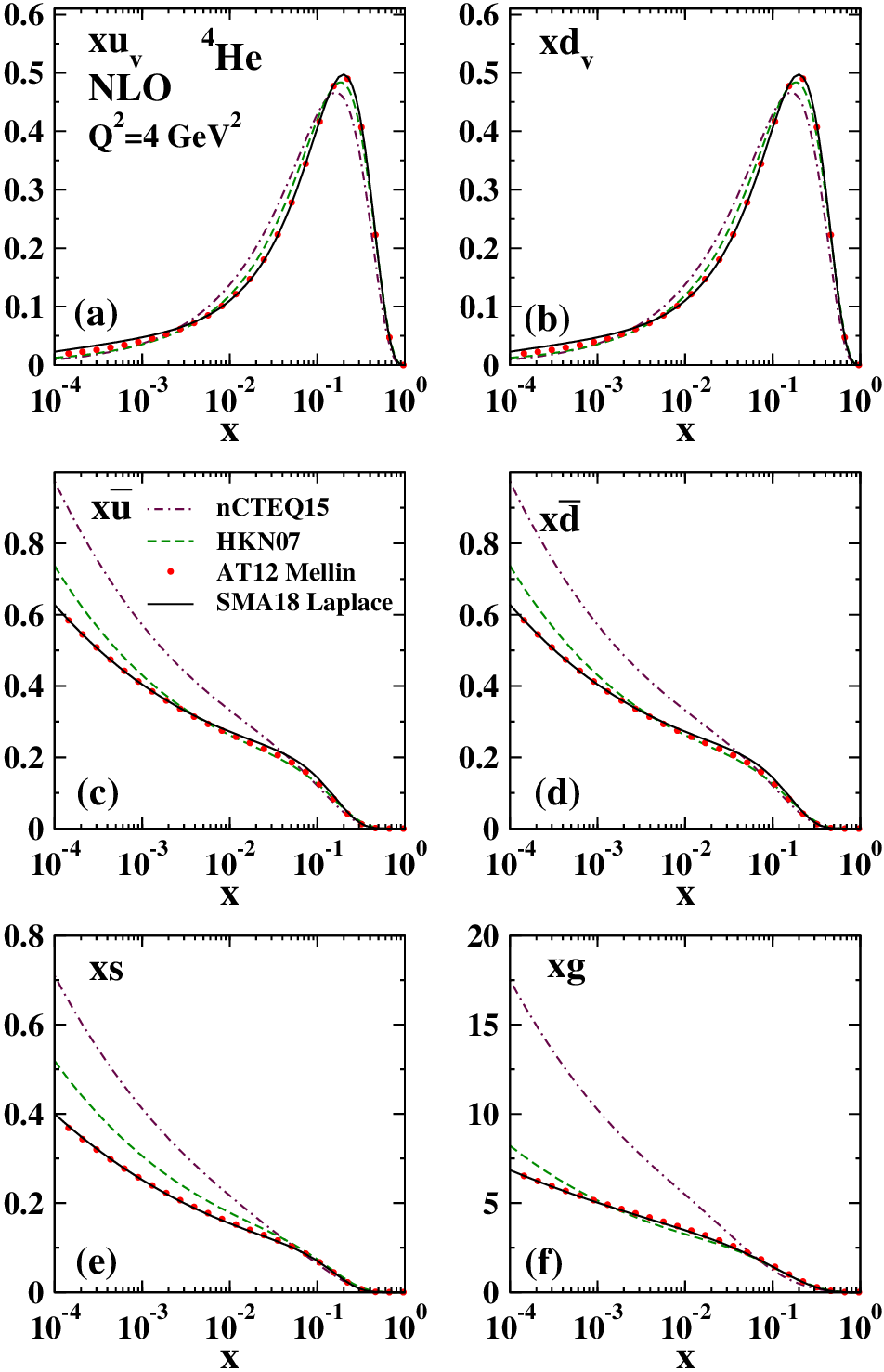}
		\caption{(Color online) Parton distribution function for $^4$He at $Q^2=4$\;GeV$^2$, resulted from  the Laplace transform, represented by the solid line (SMA18 Laplace) which has been  compared with the {\tt AT12}~\cite{AtashbarTehrani:2012xh}, {\tt HKN07}~\cite{Hirai:2007sx} and {\tt nCTEQ15}~\cite{Kovarik:2015cma} models. The plots for different types of PDFs are specified by sub-level (a) to (f). }\label{fig4:partonHe}
	\end{center}
\end{figure}
\begin{figure}[htbp]
	\begin{center}
		\includegraphics[clip,width=0.45\textwidth]{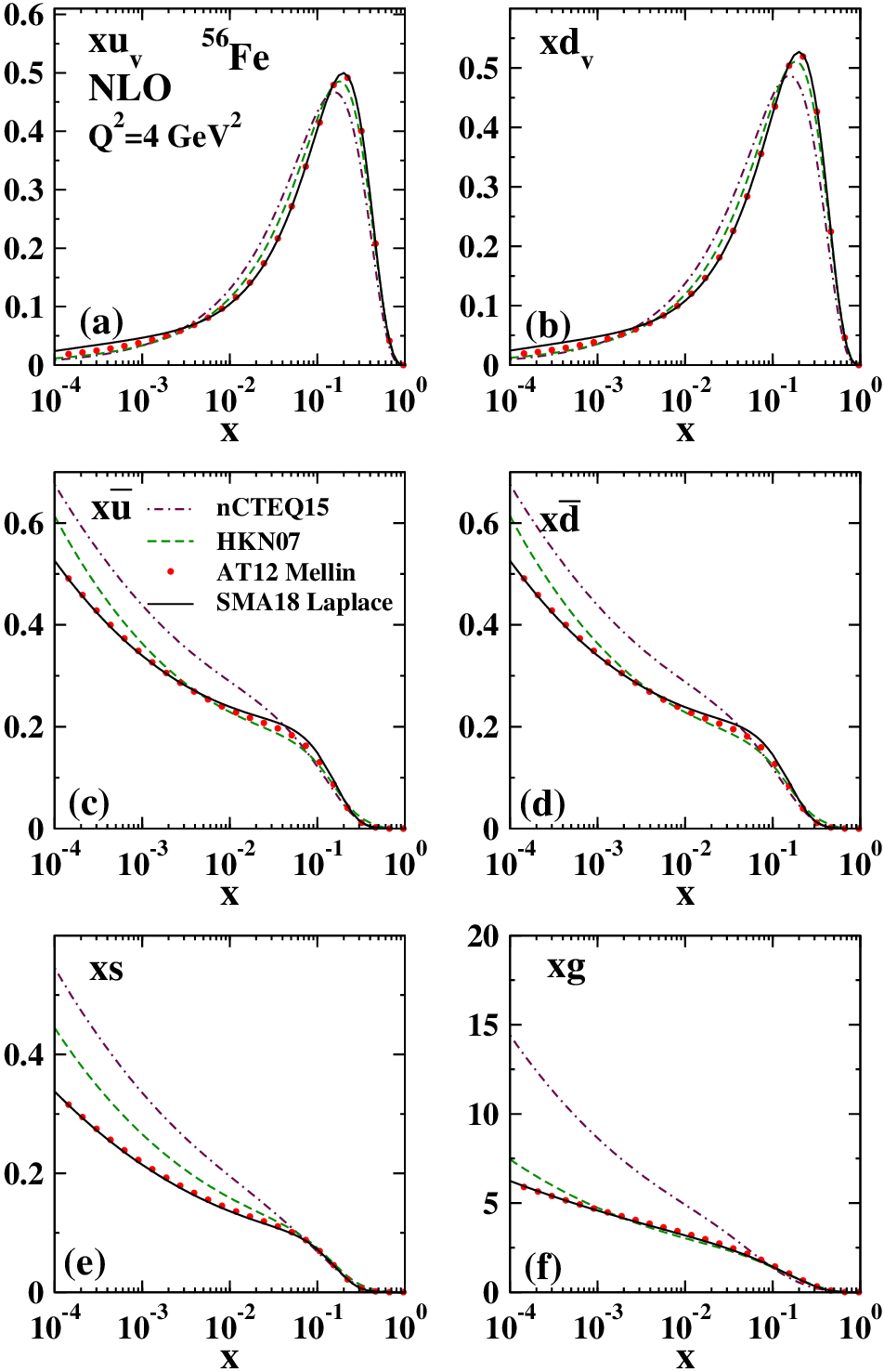}
		\caption{(Color online)  Parton distribution function for $^{56}$Fe at $Q^2=4$\; GeV$^2$ , resulted from  the Laplace transform, represented by the solid line (SMA18 Laplace) which has been  compared with the {\tt AT12}~\cite{AtashbarTehrani:2012xh}, {\tt HKN07}~\cite{Hirai:2007sx} and {\tt nCTEQ15}~\cite{Kovarik:2015cma} models. The plots for different types of PDFs are specified by sub-level (a) to (f).}\label{fig5:partonFe}
	\end{center}
\end{figure}

Now if we intend to get the solution for the nonsinglet part, $F_{\rm NS}(x,Q^2)$, its NLO contribution
can be written as :
\begin{eqnarray}\label{eq:nonsinglet}
	\frac{4\pi}{\alpha_s(Q^2)} && \frac{\partial F_{\rm NS}}{\partial ln Q^2} (x,Q^2) \nonumber \\
	&& = F_{NS}\otimes \left(p_{qq}^{\rm LO, NS} + \frac{\alpha_s(Q^2)}{4\pi}p_{qq}^{\rm NLO, NS}\right)(x,Q^2)\,. \nonumber \\
\end{eqnarray}
The first order differential equations in Laplace $s$-space for the non-singlet distribution and
{{}in terms of} the $\tau$ variable, can be obtained as in
~\cite{Block:2010du,Khanpour:2016uxh,Block:2011xb}:

$f_{\rm NS}(s,\tau)$,
\begin{eqnarray}\label{eq:nonsinglet-laplace-space}
	\frac{\partial f_{\rm NS}}{\partial \tau}(s,\tau) = \left ( \Phi_{\rm NS}^{\rm LO} + \frac{\alpha_s(\tau)}{4\pi}\Phi_{\rm NS,qq}^{\rm NLO} \right ) f_{\rm NS}(s,\tau)\,.
\end{eqnarray}
The solution of the above equation {{}would simply be}
\begin{eqnarray}\label{eq:solve-nonsinglet}
	f_{\rm NS}(s,\tau) = e^{\tau\Phi_{\rm NS}(s)}f^0_{\rm NS}(s)\;.
\end{eqnarray}
Here $\Phi_{\rm NS}(s)$ {{}includes the NLO contribution of the splitting functions
in $s$ space} such that
\begin{eqnarray}\label{eq:fi-nonsinglet}
	\Phi_{\rm NS}(s)\equiv\Phi_{\rm NS}^{\rm LO}(s)+\frac{\tau_2}{\tau}\Phi_{\rm NS,qq}^{\rm NLO}(s)\,.
\end{eqnarray}
{{}Evaluation of
$\Phi_{\rm NS, qq}^{\rm NLO}(s) = {\cal L} \left[e^{-\nu}p_{qq}^{\rm NLO, NS}(e^{-\nu}); s \right]$ is
too lengthy but straightforward and its analytical result in the transformed Laplace $s$ space at NLO
approximation is given in Appendix~{\bf A} of Ref.~\cite{Khanpour:2016uxh}.}

To amend the notations which are used in the article, it should be noted that at the leading order
approximation, {{}$Q^2$ dependence} of the evolution equation is in fact represented by $\tau$ variable and at the NLO approximation, by $\tau_2$; the {{}former is defined in Refs.}~\cite{Block:2010du,Khanpour:2016uxh,Block:2011xb,AtashbarTehrani:2013qea},
\begin{eqnarray}\label{eq:tau2}
	\tau_2 \equiv \frac{1}{4\pi} \int_0^{\tau} \alpha(\tau^{\prime}) d\tau^{\prime} \ = \ (\frac{1}{4\pi})^2 \int_{Q_0^2}^{Q^2} \alpha_s^2 (Q^{\prime 2}) \ d \ln Q^{\prime 2}\,. \nonumber \\
\end{eqnarray}

We should use the {{} variable $\tau_2$ since the current analysis is done at NLO approximation
but for simplicity in the notation, $\tau$ will be used in the remainder of the paper in its place.} It should
be finally noted that the NLO expansion parameter, $a_1$, in the iterative solution of Eq.(\ref{eq:atau})
is quite small. {{}For instance $a_1=0.025$ for $M^2_c<Q^2 \leq M^2_b GeV^2$ while
$b_1=10.7$ and $a_1=0.017$ with $b_1=8.63$ for $M^2_b<Q^2\leq 10^5 GeV^2$. Furthermore,
 $a_0=0.025$ and is constant over the whole range of $Q^2$ scale ~[\ref{eq:spliLOandNLO}].}

\section{Theoretical formalism for the EMC effect}\label{Sec3:emc-effect}
To calculate the parton distribution in nuclear media, {{}we would need to have
the parton distributions for a free proton. To achieve this, it is required to use a set of PDFs
at the input scale $Q_0^2=2\; GeV^2$ which are depicting in Fig. \ref{fig1:freepartonproton}
and have the following standard parametrization, as in Ref.~\cite{Khanpour:2012tk}}:

\begin{eqnarray}\label{eq:freeproton}
  xu_v&=&0.37328x^{0.32182}(1-x)^{3.59165} \nonumber\\
  &&(1+3.62456x^{0.50629}+21.31705x),\nonumber\\
  xd_v&=&0.51354x^{0.39354}(1-x)^{5.03622} \nonumber\\
  &&(1-1.26057x^{0.47037}+15.98368x),\nonumber\\
  2x(\overline{d}+\overline{u})&=&0.29795x^{-0.2052}(1 - x)^{9.06901}\nonumber\\
  &&(1+0.93542x^{0.33012}+14.46062x),\nonumber\\
  x(\overline{d}-\overline{u})&=&9.49265x^{1.33727}(1-x)^{18.559}\nonumber\\
  &&(1-7.82741x^{0.54431}+20.60532x),\nonumber\\
  xs&=&0.03724x^{-0.2052}(1-x)^{9.06901}\nonumber\\
  &&(1+0.93542x^{0.33012}+14.46062x),\nonumber\\
  xg&=&3.60703x^{0.062467}(1-x)^{6.75001}\nonumber\\
  &&(1+3.91106x^2-0.813601x).
\end{eqnarray}

On the other hand, using a number of parameters, the nPDFs are specified at a fixed $Q^2$ which is
usually taken as $Q_0^2$. There is a relation between the nPDFs and the PDFs in free proton
in which the PDFs are multiplied by a weight function  $w_i$ such that  \cite{Hirai:2001np}:

\begin{equation}\label{eq:weightfunction1}
f_i^A(x,Q_0^2)=w_i(x,A,Z)f_i(x,Q_0^2)
\end{equation}

Using a $\chi^2$ analysis procedure, the parameters in the weight function which are dependent on $x$,
$A$ ({{}atomic mass}), and $Z$ (atomic number), can be obtained.

{{}The following functional forms would be assumed for the weight function in
Eq.~(\ref{eq:weightfunction1}) which are based on the analysis in Refs. \cite{Hirai:2001np,Hirai:2004wq,Hirai:2007sx,Tehrani:2004hp,Tehrani:2006gy,Tehrani:2007hu,AtashbarTehrani:2012xh,Khanpour:2016pph}}:
\begin{eqnarray}\label{eq:weightfunction2}
  w_i &=& 1+\left(1-\frac{1}{A^{\alpha_i}}\right)\nonumber \\
   && \times\frac{a_i(A,Z)+b_i(A)x+c_i(A)x^2+d_i(A)x^3}{(1-x)^{\beta_i}}
\end{eqnarray}

The following nPDFs can be obtained, considering the weight function in Eq.~(\ref{eq:weightfunction2}) which is combined with PDFs in Eq.~(\ref{eq:freeproton}):

\begin{eqnarray}
  u_v^A(x,Q_0^2) &=&w_{u_v}(x,A,Z)\frac{Zu_v(x,Q_0^2)+Nd_v(x,Q_0^2)}{A},\nonumber \\
  d_v^A(x,Q_0^2) &=&w_{d_v}(x,A,Z)\frac{Zd_v(x,Q_0^2)+Nu_v(x,Q_0^2)}{A},\nonumber\\
  \overline{u}^A(x,Q_0^2) &=&w_{\overline{q}}(x,A,Z)\frac{Z\overline{u}(x,Q_0^2)+N\overline{d}(x,Q_0^2)}{A},\nonumber\\
  \overline{d}^A(x,Q_0^2) &=&w_{\overline{q}}(x,A,Z)\frac{Z\overline{d}(x,Q_0^2)+N\overline{u}(x,Q_0^2)}{A},\nonumber\\
  s^A(x,Q_0^2) &=&w_{\overline{q}}(x,A,Z)s(x,Q_0^2),\nonumber\\
  g^A(x,Q_0^2) &=&w_{g}(x,A,Z)g(x,Q_0^2).\label{npdf}
\end{eqnarray}

The $Z$ term and the $N (= A - Z)$ term in the above equations are indicating the atomic number
(the number of protons) and the number of neutrons in the nuclei respectively.
Here the SU(3) symmetry  is not assumed.

For the case of isoscalar nuclei in which the number of protons and neutrons in a nucleus
are equal to each other, valence quarks  as well as anti-quark would have similar distributions.
{{}But since in heavy nuclei the number of the  neutrons  is larger than the number of protons
$(N>Z)$, as can be seen in Eq.(\ref{npdf}), the distribution of down valence quarks would be greater
than that of up valence quarks. Following that, it can be seen, as well, that in this type of the nuclei, antiquark
distributions $(\overline{u}^A,\overline{d}^A,\overline{s}^A)$  would not be equal to each other  ~\cite{Kumano:1997cy,Garvey:2001yq}.}

As in Ref.~\cite{Sick:1992pw}, {{}$\alpha_i$ is taken to have the value $1/3$ in
Eq.~(\ref{eq:weightfunction2})}. {{}Also it should be noted that, there exist three constraints on the parameters
in the equation, considering nuclear volume and surface contributions. These  constraints are related to
the nuclear charge $Z$, baryon number ( atomic number) $A$ and momentum conservation
~\cite{Hirai:2001np,Hirai:2004wq,AtashbarTehrani:2012xh,Frankfurt:1990xz}, which
can be written in the Laplace space as follows:}
\begin{eqnarray}
  Z &=&\frac{A}{3}{\cal L}[2e^{-v}u_v^A(e^{-v},Q_0^2)-e^{-v}d_v^A(e^{-v},Q_0^2);s=0],\nonumber \\
  A &=& \frac{A}{3}{\cal L}[e^{-v}u_v^A(e^{-v},Q_0^2)+e^{-v}d_v^A(e^{-v},Q_0^2);s=0],\nonumber \\
  A &=& A{\cal L}\{e^{-v}[u_v^A+d_v^A+2(\overline{u}^A+\overline{d}^A+s^A)+g^A](e^{-v},Q_0^2);s=1\}.\nonumber\\
\end{eqnarray}

\begin{widetext}
In order to be able to do the required calculations for iron (Fe), calcium (Ca), carbon (C), helium  (He)
and deuterium (D) nuclei, we need {{}the relevant weight functions which are presented in the following relations in which the effects of shadowing, anti-shadowing, fermi motion and the EMC regions  are included ~\cite{AtashbarTehrani:2012xh}:}
\begin{eqnarray}
  w_{u_v}^{Fe} &=& 1+\left(1-\frac{1}{56^{1/3}}\right)\frac{-0.0979153+2.08684x-6.91749x^2+5.50217x^3}{(1-x)^{0.4}} \nonumber\\
  w_{d_v}^{Fe} &=& 1+\left(1-\frac{1}{56^{1/3}}\right)\frac{-0.0980722+2.08684x-6.91749x^2+5.50217x^3}{(1-x)^{0.4}} \nonumber\\
  w_{\overline{q}}^{Fe} &=& 1+\left(1-\frac{1}{56^{1/3}}\right)\frac{-0.344557+7.71619x-45.8738x^2+66.9498x^3}{(1-x)^{0.1}} \nonumber\\
  w_{g}^{Fe} &=& 1+\left(1-\frac{1}{56^{1/3}}\right)\frac{-0.305125+2.59586x+0.369233x^3}{(1-x)^{0.1}}\\
  w_{u_v}^{Ca} &=& 1+\left(1-\frac{1}{40^{1/3}}\right)\frac{-0.0972234+2.08106x-6.90323x^2+5.47457x^3}{(1-x)^{0.4}} \nonumber\\
  w_{d_v}^{Ca} &=& 1+\left(1-\frac{1}{40^{1/3}}\right)\frac{-0.0972234+2.08106x-6.90323x^2+5.47457x^3}{(1-x)^{0.4}} \nonumber\\
  w_{\overline{q}}^{Ca} &=& 1+\left(1-\frac{1}{40^{1/3}}\right)\frac{-0.325852+7.19981x-42.7529x^2+60.2908x^3}{(1-x)^{0.1}} \nonumber\\
  w_{g}^{Ca} &=& 1+\left(1-\frac{1}{40^{1/3}}\right)\frac{-0.300195+2.59586x+0.369233x^3}{(1-x)^{0.1}}\\
   w_{u_v}^{C} &=& 1+\left(1-\frac{1}{12^{1/3}}\right)\frac{-0.0944919+2.0605x-6.85243x^2+5.37693x^3}{(1-x)^{0.4}} \nonumber\\
  w_{d_v}^{C} &=& 1+\left(1-\frac{1}{12^{1/3}}\right)\frac{-0.0944919+2.0605x-6.85243x^2+5.37693x^3}{(1-x)^{0.4}} \nonumber\\
  w_{\overline{q}}^{C} &=& 1+\left(1-\frac{1}{12^{1/3}}\right)\frac{-0.266859+5.61929x-33.2257x^2+41.4429x^3}{(1-x)^{0.1}} \nonumber\\
  w_{g}^{C} &=& 1+\left(1-\frac{1}{12^{1/3}}\right)\frac{-0.288775+2.59586x+0.369233x^3}{(1-x)^{0.1}}\\
  w_{u_v}^{He} &=& 1+\left(1-\frac{1}{4^{1/3}}\right)\frac{-0.0920426+2.04191x-6.8064x^2+5.28935x^3}{(1-x)^{0.4}} \nonumber\\
  w_{d_v}^{He} &=& 1+\left(1-\frac{1}{4^{1/3}}\right)\frac{-0.0920426+2.04191x-6.8064x^2+5.28935x^3}{(1-x)^{0.4}} \nonumber\\
  w_{\overline{q}}^{He} &=& 1+\left(1-\frac{1}{4^{1/3}}\right)\frac{-0.2224+4.48189x-26.3976x^2+29.4371x^3}{(1-x)^{0.1}} \nonumber\\
  w_{g}^{He} &=& 1+\left(1-\frac{1}{4^{1/3}}\right)\frac{-0.284205+2.59586x+0.369233x^3}{(1-x)^{0.1}}\allowdisplaybreaks[1]\\
  w_{u_v}^{D} &=& 1+\left(1-\frac{1}{2^{1/3}}\right)\frac{-0.0905182+2.03027x-6.77752x^2+5.23484x^3}{(1-x)^{0.4}} \nonumber\\
  w_{d_v}^{D} &=& 1+\left(1-\frac{1}{2^{1/3}}\right)\frac{-0.0905182+2.03027x-6.77752x^2+5.23484x^3}{(1-x)^{0.4}} \nonumber\\
  w_{\overline{q}}^{D} &=& 1+\left(1-\frac{1}{2^{1/3}}\right)\frac{-0.198243+3.8859x-22.8312x^2+23.7229x^3}{(1-x)^{0.1}} \nonumber\\
  w_{g}^{D} &=& 1+\left(1-\frac{1}{2^{1/3}}\right)\frac{-0.283108+2.59586x+0.369233x^3}{(1-x)^{0.1}}
   \end{eqnarray}

\end{widetext}

In Fig.~\ref{fig2:weightfunction}, the weight functions for the Fe, Ca, C, He, and D nuclei are depicting
at {{}the} initial scale $Q_0^2=2$ \;GeV$^2$ and in  Fig.~\ref{fig3:FeQ2}, the parton distribution
functions inside $^{56}Fe$ nucleus at $Q_0^2=2$\;GeV$^2$ are {{}presented}.
\begin{figure}[htbp]
	\begin{center}
		\includegraphics[clip,width=0.5\textwidth]{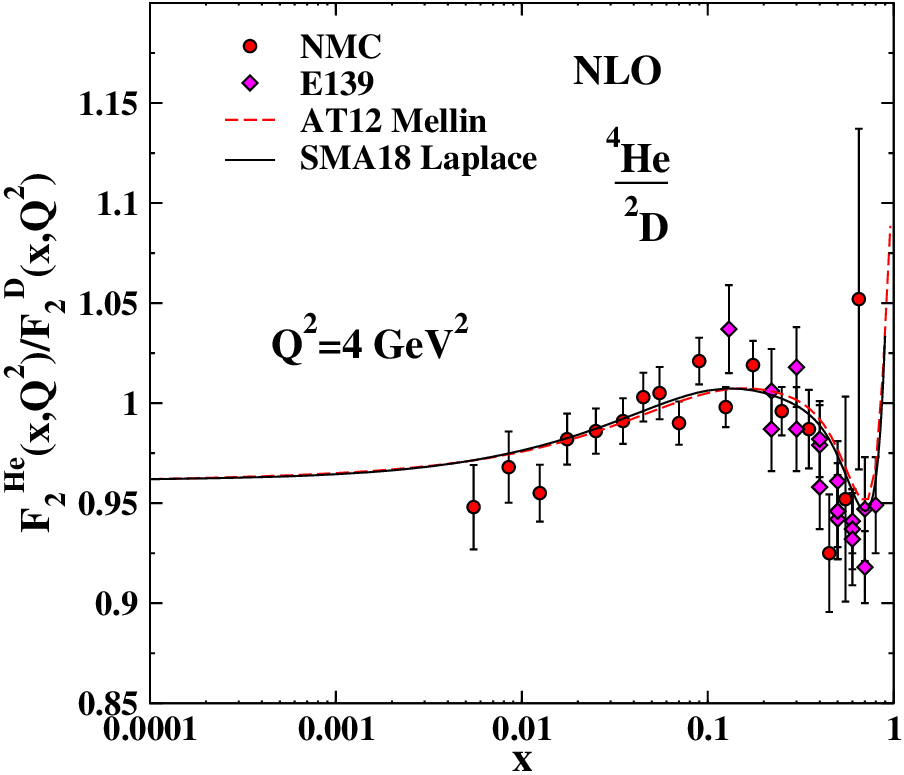}
		\caption{(Color online) EMC effect for $^{4}$He/$^{2}$D at $Q^2=4$\;GeV$^2$, resulted from the Laplace transform , represented by solid line (SMA18 Laplace) which has been  compared with the {\tt AT12} model ~\cite{AtashbarTehrani:2012xh}  and DIS data of NMC~\cite{Amaudruz:1995tq} and E139~\cite{Gomez:1993ri} Collaborations.}\label{fig6:emcHe}
	\end{center}
\end{figure}

\begin{figure}[htbp]
	\begin{center}
		\includegraphics[clip,width=0.5\textwidth]{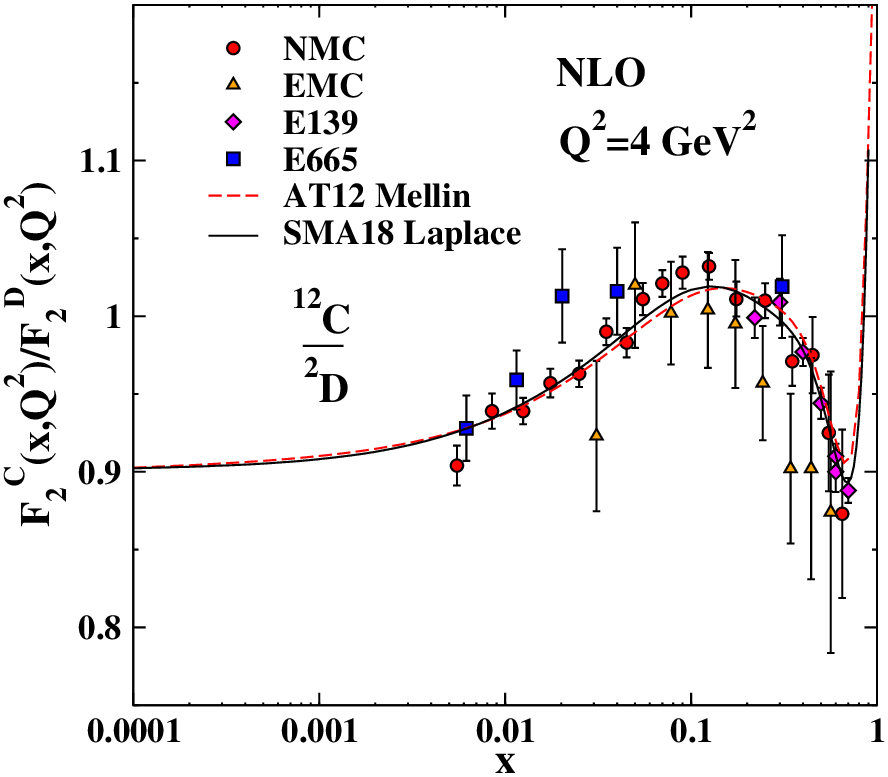}
		\caption{(Color online) EMC effect for $^{12}$C/$^{2}$D at $Q^2=4$\;GeV$^2$, resulted from the Laplace transform, , represented by solid line (SMA18 Laplace) which has been  compared with the {\tt AT12}  model~\cite{AtashbarTehrani:2012xh}   and DIS data of NMC~\cite{Arneodo:1995cs}, EMC~\cite{Ashman:1988bf}, E139~\cite{Gomez:1993ri} and E665~\cite{Adams:1995is} Collaborations.}\label{fig7:emcC}
	\end{center}
\end{figure}

\begin{figure}[htbp]
	\begin{center}
		\includegraphics[clip,width=0.5\textwidth]{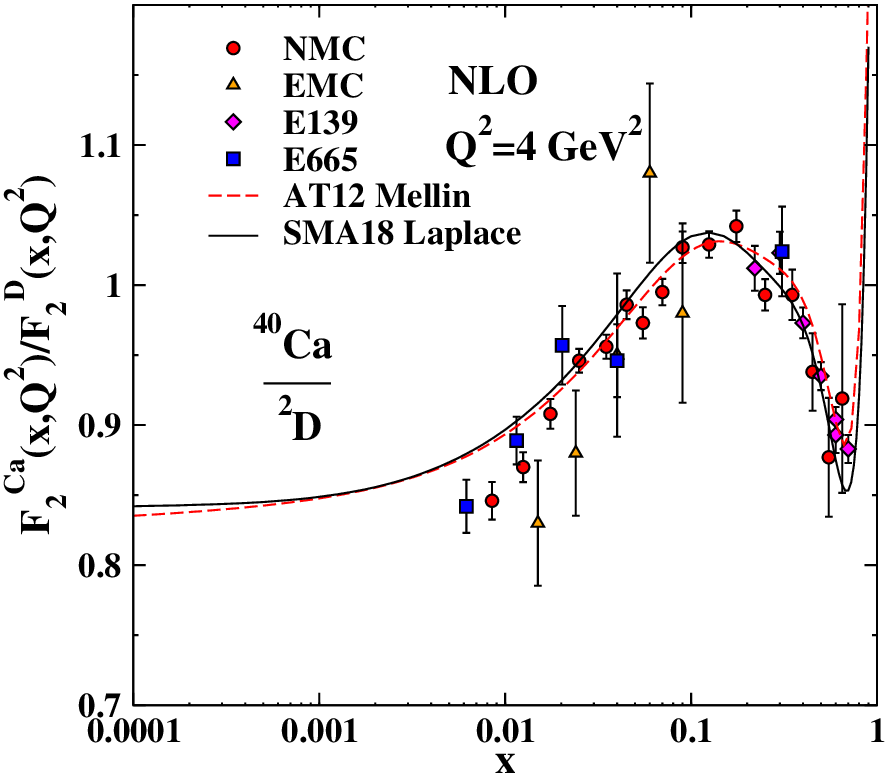}
		\caption{(Color online) EMC effect for $^{40}$Ca/$^{2}$D at $Q^2=4$\;GeV$^2$ resulted from the Laplace transform, represented by solid line (SMA18 Laplace) which has been  compared with the {\tt AT12} model~\cite{AtashbarTehrani:2012xh} and DIS data of NMC~\cite{Arneodo:1995cs}, EMC~\cite{Ashman:1988bf}, E139~\cite{Gomez:1993ri} and E665~\cite{Adams:1995is} Collaborations.}\label{fig8:emcCa}
	\end{center}
\end{figure}
\begin{figure}[htbp]
	\begin{center}
		\includegraphics[clip,width=0.5\textwidth]{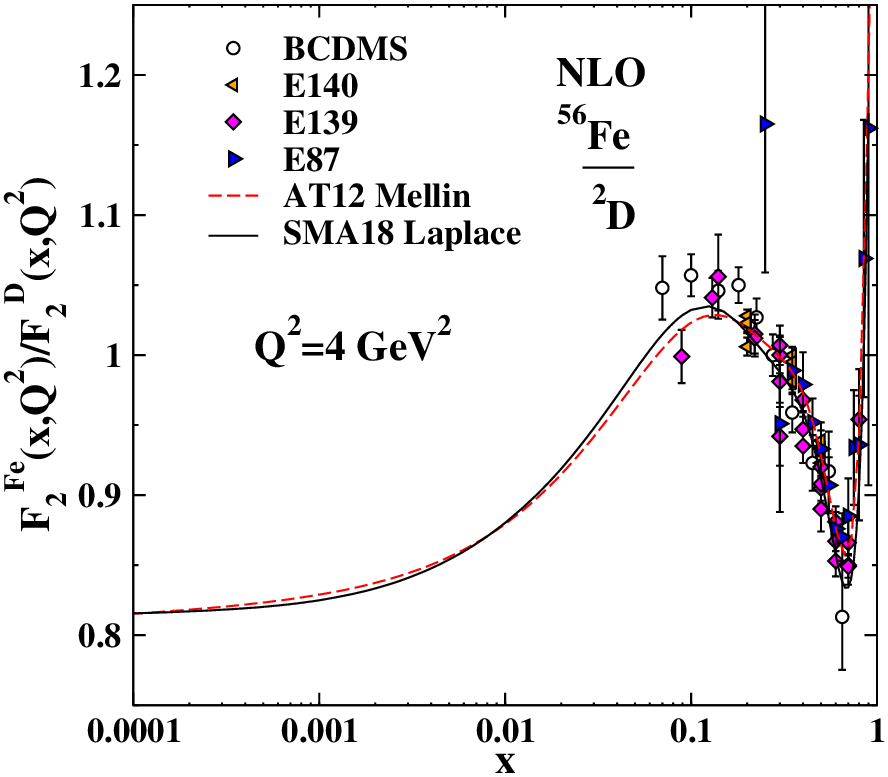}
		\caption{(Color online) EMC effect for $^{56}$Fe/$^{2}$D at $Q^2=4$\;GeV$^2$ resulted from the Laplace transform, represented by a solid line (SMA18 Laplace) which has been  compared with the {\tt AT12} model~\cite{AtashbarTehrani:2012xh}  and DIS data of BCDMS~\cite{Benvenuti:1987az}, E140~\cite{Dasu:1988ru}, E139~\cite{Gomez:1993ri} and finally E87~\cite{Bodek:1983qn} Collaborations.}\label{fig9:emcFe}
	\end{center}
\end{figure}

\begin{figure}[htbp]
	\begin{center}
		\includegraphics[clip,width=0.5\textwidth]{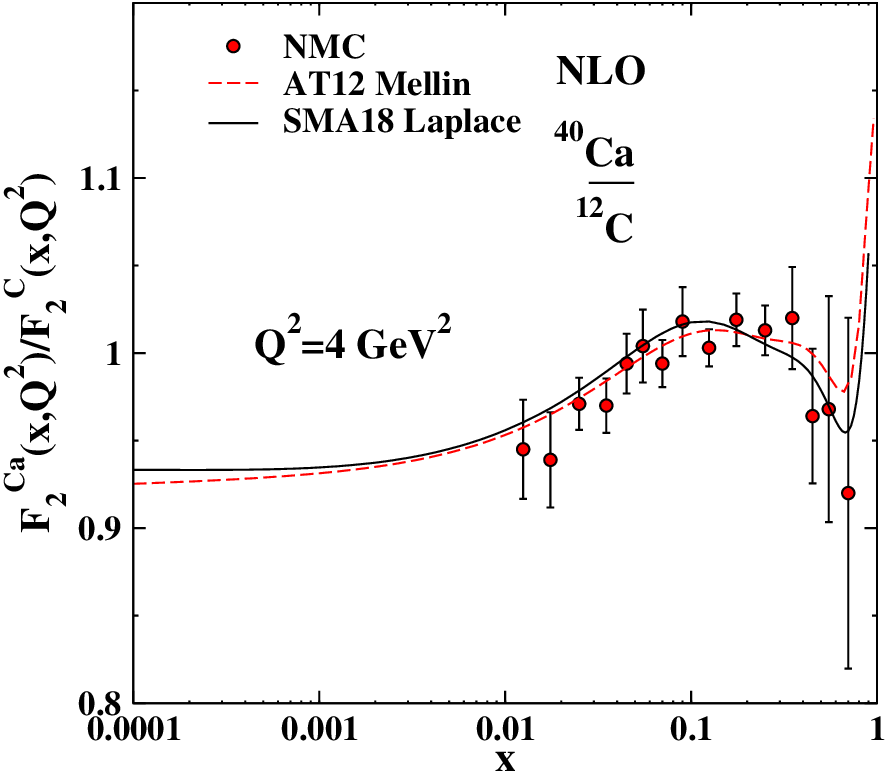}
		\caption{(Color online) EMC effect for $^{40}Ca/^{12}C$ at $Q^2=4\;GeV^2$ resulted from the Laplace transform, represented by solid line (SMA18 Laplace) which has been  compared with the {\tt AT12} model~\cite{AtashbarTehrani:2012xh} and DIS data in nuclear reaction NMC~\cite{Amaudruz:1995tq}.}\label{fig10:emcCatoC}
	\end{center}
\end{figure}

\begin{figure}[htbp]
	\begin{center}
		\includegraphics[clip,width=0.5\textwidth]{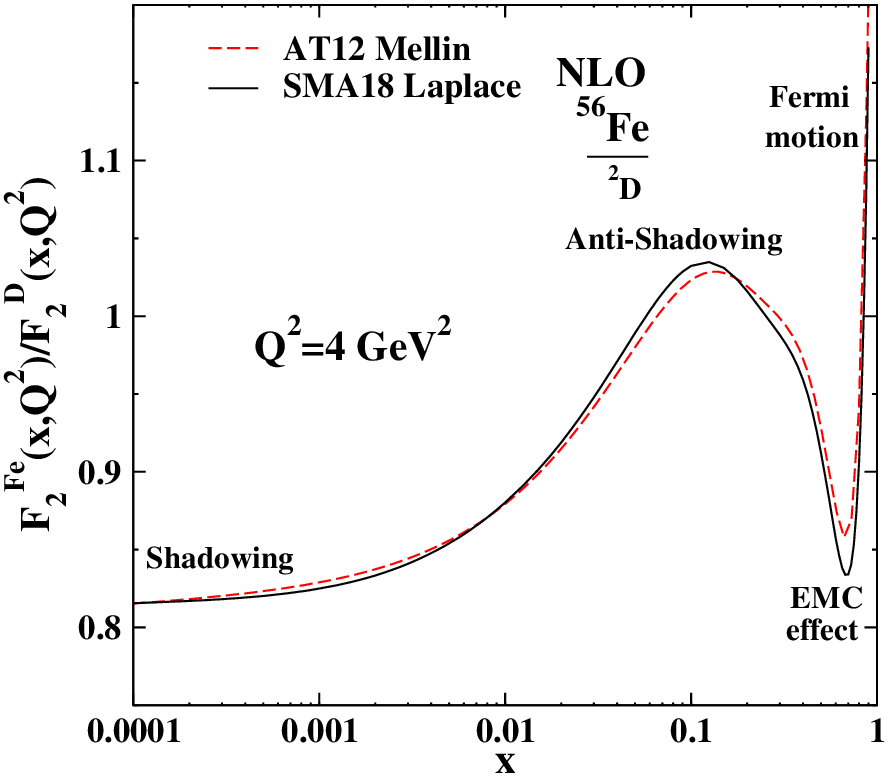}
		\caption{(Color online) EMC effect for $^{56}$Fe/$^{2}$D at $Q^2=4$\;GeV$^2$ resulted from the Laplace transform, represented by a solid line (SMA18 Laplace) which has been  compared with the {\tt AT12} model~\cite{AtashbarTehrani:2012xh} while the { anti-shadowing, shadowing, EMC and fermi motion regions are included. }}\label{fig11:emcFetoD}
	\end{center}
\end{figure}

\section{Nuclear structure function $F_2^A(x, Q^2)$ in the Laplace space}\label{Sec4:Nuclear-structure-function}
Based on the Laplace transform technique, we perform here an analytical {{}calculation of} the
nuclear structure  function $F_2^A(x, Q^2)$ at NLO approximation. {{}We should first extract
the nucleon structure function, using the singlet, gluon and non-singlet parton distributions which were
obtained in the previous sections.}
The nuclear structure function $F_2^A(x, Q^2)$ in Laplace $s$ space, up to the next-to-leading order approximation, can be written as
\begin{eqnarray}\label{eq:F2ALaplacespacelight}
	{\cal F}_2^{\rm A} (s, \tau) =  {\cal F}_2^{\rm S} (s, \tau) + {\cal F}_2^G(s, \tau) + {\cal F}_2^{\rm NS}(s, \tau) \,,
\end{eqnarray}
where the flavour singlet ${\cal F}_2^{\rm S}$ and gluon ${\cal F}_2^G$ contribution reads
\begin{eqnarray}\label{eq:F2pLaplacespace-singlet}
	{\cal F}_2^{\rm S}(s,\tau) & = & \left(\frac{4}{9}2\bar{u}^A(s,\tau)  +  \frac{1}{9}2\bar{d}^A(s,\tau)  + \frac{1}{9}2\bar{s}^A(s,\tau)\right)  \nonumber  \\
	&& \times \left(1 + \frac{\tau}{4\pi}C^{(1)}_q(s) \right) \,,
\end{eqnarray}
\begin{eqnarray}\label{eq:F2pLaplacespace-gluon}
	{\cal F}_2^{\rm G}(s,\tau)& = &  \frac{2}{9} g^A(s,\tau)\left(\frac{\tau}{4\pi}C^{(1)}_g(s)\right)\,.
\end{eqnarray}
Finally, the non-singlet contribution for three active (light) flavours is given by
\begin{eqnarray}\label{eq:F2pLaplacespace-nonsinglet}
	{\cal F}_2^{\rm NS}(s,\tau) &=&\left(\frac{4}{9} u_v^A(s,\tau) + \frac{1}{9}d_v^A(s,\tau)\right)\left(1+ \frac{\tau}{4\pi}C^{(1)}_q(s)\right)\;, \nonumber\\
\end{eqnarray}
where {{}$C_q^{(1)}(s)$ and $C_g^{(1)}(s)$ represent Wilson coefficients functions at NLO and
can be derived in Laplace $s$ space by $c_q(s) = {\cal L} [e^{-\nu} c_q(e^{-\nu}); s]$ and
$c_g(s) = {\cal L} [e^{-\nu} c_g(e^{-\nu}); s]$. Explicit expressions for the corresponding Wilson coefficients functions are as follow~\cite{Khanpour:2016uxh}:}
\begin{widetext}
\begin{eqnarray}
	&& C_q^{(1)}(s)= \nonumber\\
	&& C_F \left(-9-\frac{2 \pi ^2}{3}-\frac{2}{(1+s)^2}+\frac{6}{1+s}-\frac{2}{(2+s)^2}+\frac{4}{2+s}+\right. \nonumber\\
	&&3 \left(\gamma _E+\psi (s+1)\right)+ \frac{2 \left(\gamma _E+\psi (s+2)\right)}{1+s}+\frac{2 \left(\gamma _E+\psi (s+3)\right)}{2+s}+ \nonumber\\
	&& \left.\frac{1}{3} \left(\pi ^2+6 \left(\gamma _E+\psi (s+1)\right){}^2-6 \psi '(s+1)\right)+4 \psi '(s+1)\right)\,,\nonumber\\
\end{eqnarray}
\begin{eqnarray}
	&& C_g^{(1)}(s) =  \nonumber\\
	&& f \left(\frac{2}{(1+s)^2}-\frac{2}{1+s}-\frac{4}{(2+s)^2}+\frac{16}{2+s}+\frac{4}{(3+s)^2}-\frac{16}{3+s}-\right. \nonumber\\
	&& \left.\frac{2 \left(\gamma _E+\psi (s+2)\right)}{1+s}+\frac{4 \left(\gamma _E+\psi (s+3)\right)}{2+s}-\frac{4 \left(\gamma _E+\psi (s+4)\right)}{3+s}\right)\,.\nonumber\\
\end{eqnarray}
\end{widetext}

As before, {{} $Q^2$ dependence of the nuclear structure function} in
Eq.~(\ref{eq:F2ALaplacespacelight}) is  given again by $\tau(Q^2, Q_0^2) \equiv {1 \over 4\pi} \int_{Q^2_0}^{Q^2} \alpha_s({Q^{\prime}}^2) d \, \ln {Q^{\prime}}^2$.
Using the inverse Laplace transform and the appropriate change of variables \cite{Khanpour:2016uxh},
the desired solution for the nuclear structure function in Bjorken $x$ space, $F_2^{\rm A}(x,Q^2)$, can
be readily obtained.

\section{Laplace transformation technique and the EMC results}\label{Sec5:Results}
Based on the analytical solution for the DGLAP evolution equations,  using the Laplace transformation
technique, we shall first present in this section our results that have been obtained for the parton
distribution functions {{}after which} the nuclear structure function ratio
$F_2^{A^{\prime}}(x,Q^2)/F_2^A(x,Q^2)$ would be presented. {{}Figure~
\ref{fig2:weightfunction} illustrates the weight function for $^{56}$Fe, $^{40}$Ca, $^{12}$C, $^4$He
and $^2$D nuclei.} {In Fig.~\ref{fig3:FeQ2}, we depict parton distribution functions inside
$^{56}Fe$  nucleus at $Q_0^2=2$\;GeV$^2$  while according to Eq. ({\ref{npdf}}) {{}
SU(3) symmetry breaking is supposed to be in place and hence the down anti-quark distribution is
assumed to be larger than the up anti-quark distribution where additionally we see as well that down valence
quark distribution is greater than up valence distribution}. Based on  Eq. \eqref{eq:solve-nonsinglet},
{{}the required calculations could be performed to obtain, at the NLO approximation,
the valence quark distributions, $x u_v^A (x,Q^2)$ and $x d_v^A (x,Q^2)$.
In Figs.~\ref{fig4:partonHe},and~\ref{fig5:partonFe}, these distributions have been presented alongside
other parton distributions, including the anti-quarks and gluon distribution functions  for $^4$He and
$^{56}$Fe nuclei  at $Q^2=4$\;GeV$^2$ in Laplace $s$-space. Furthermore, they have been compared with
{\tt AT12}~\cite{AtashbarTehrani:2012xh}, {\tt HKN07}~\cite{Hirai:2007sx} and {\tt nCTEQ15}~
\cite{Kovarik:2015cma} models.} {{}As can be seen, a good agreement does exist between the
presented results and the results obtained from the other models.
The solid line represents our solution, resulting from the Laplace transform technique and the red
circles represent the parton quark distributions from the {\tt AT12} model.}

In Fig.~\ref{fig6:emcHe}, {{}the EMC effect has been demonstrated for $^4$He nucleus in Laplace s space and Mellin space~\cite{AtashbarTehrani:2012xh} at $Q^2=4$\; GeV$^2$ and compared with DIS data in nuclear reactions from NMC~\cite{Amaudruz:1995tq} and E139~\cite{Gomez:1993ri}
Collaborations.}
{{}The EMC effect for $^{12}C$ nucleus in Laplace s-space and Mellin momment space~
\cite{AtashbarTehrani:2012xh} at $Q^2=4$\;GeV$^2$ has been shown in Fig.~\ref{fig7:emcC} and they
have been compared with the results from the DIS data in nuclear reactions NMC~\cite{Arneodo:1995cs}
as well as  EMC~\cite{Ashman:1988bf}, E139~\cite{Gomez:1993ri} and E665~\cite{Adams:1995is}
collaborations.}
{{} The corresponding results for the EMC effect in $^{40}$Ca structure function in Laplace $s$-
space and Mellin moment space~\cite{AtashbarTehrani:2012xh} at $Q^2=4$\; GeV$^2$ have been
presented inn Fig.~\ref{fig8:emcCa} and compared with the results from DIS data in
nuclear reactions by NMC~\cite{Amaudruz:1995tq} and EMC~\cite{Arneodo:1989sy}, E139~
\cite{Gomez:1993ri}, and E665~\cite{Adams:1995is} Collaborations.}

It is seen that our analytical solutions based on the inverse Laplace transform technique at the NLO
approximation for the nuclear structure function over a wide range of $x$ and $Q^2$ values,
correspond well with the experimental data and the {{}{\tt AT12} model}.
One can conclude that, in spite of small disagreements for the parton densities, we find a satisfactory
agreement for {{}the nuclear structure function ratio over a wide range of $x$'s and $Q^2$'s.}
{{}The overall agreement is found to have a deviation of 1 part in 10$^5$.}
In Fig.~\ref{fig9:emcFe}, we present the EMC effect for $^{56}$Fe structure function in Laplace space
and Mellin space~\cite{AtashbarTehrani:2012xh} at $Q^2=4$\;GeV$^2$ and a comparison with the DIS
data in nuclear reactions BCDMS~\cite{Benvenuti:1987az}, and also E140~\cite{Dasu:1988ru}, E139~
\cite{Gomez:1993ri}, and E87~\cite{Bodek:1983qn} Collaborations is done.

Finally, in Fig.~\ref{fig10:emcCatoC} we {{}compare} the EMC effect for $^{40}$Ca/$^{12}$C
structure function in Laplace $s$-space as well as {{}in the Mellin moment space}
~\cite{AtashbarTehrani:2012xh} at $Q^2=4$\; GeV$^2$  with the DIS data from the NMC Collaboration~\cite{Amaudruz:1995tq}.
{{}In all figures, including for example Fig.~\ref{fig11:emcFetoD}, the effects of the shadowing
region and Fermi region are identical in the Mellin and Laplace spaces while the effect of the anti-shawdowing
region and EMC region in Laplace space are more acceptable and better than in the Mellin space.}

\section{Summary and conclusion}\label{Sec6:Summary}
The result for NLO decoupled analytical evolution equations for singlet $F_{\rm S}(x, Q^2)$,
gluon $G(x, Q^2)$ and non-singlet $F_{\rm NS}(x, Q^2)$ have been presented in this article,
which resulted from the solution of coupled DGLAP evolution equations in the Laplace $s$-space.
{{}Following that, we performed} the required calculations and obtained the results for valence
quark distributions $x u_v$ and $x d_v$, the anti-quark distributions
$x \overline{d}$ and  $x\overline{u}$,  the strange sea distribution $x s = x \overline{s}$
and finally the gluon distribution $x g$, using the input parton distributions at
$Q_0^2$ = 2\; GeV$^2$ for free protons which is initiated from the
{\tt KKT12} model ~\cite{Khanpour:2012tk}.
We also calculated in this work, the nuclear structure function $F_2^A(x, Q^2)$
which is a direct result from the Laplace transform technique.
To derive this structure function, the corresponding analytical solutions for singlet
$F_2^{\rm S}(x, Q^2)$, gluon $F_2^{\rm G}(x, Q^2)$ and non-singlet $F_2^{\rm NS}(x, Q^2)$
structure functions, inside the nucleus, are needed.
{{}Having} the initial distributions for singlet, gluon and non-singlet distributions at the
input scale $Q_0^2$, we could obtain the nuclear structure function at any arbitrary $Q^2$ scale.
The employed method, in our analysis, {{}creates} the possibility of obtaining a strictly
analytical solution {{}in terms of} the $x$-variable for nuclear parton densities
as well as the structure function .
As a final point, we got the general solutions such that they are in satisfactory agreement with
AT12, HKN07, and nCTEQ15  models and also with the available experimental data including those of the NMC, BCDMS, E87,
E139, E140, and E65 Collaborations.
As a further research task, it is possible to extend the calculation up to NNLO approximation
to investigate the EMC effect, using the Laplace transformation while new updated data are employed.
We hope to report on this issue in future.
\section*{Acknowledgments}
The authors are indebted to F. Olness for giving the required grid data. A. M. acknowledges Yazd University for providing facilities to do this project.

\end{document}